# Characterization of Frequency-Dependent Material Properties of Human Liver and its Pathologies Using an Impact Hammer

M. Umut Ozcan [a], Sina Ocal[a], Cagatay Basdogan [a*]

Gulen Dogusoy [b], Yaman Tokat [c]

[a]College of Engineering, Koc University, Istanbul, Turkey

[b]Department of Pathology, Cerrahpasa Medical School, Istanbul University, Istanbul, Turkey

[c] Liver Transplant Center, Florence Nightingale Hospital, Istanbul, Turkey

---

[*] Corresponding author:
Prof. Cagatay Basdogan
College of Engineering
Koc University
Rumelifeneri yolu, Istanbul, 34450, Turkey
Tel: 90+ 212-338-1721
Fax: 90+ 212-338-1548
e-mail: cbasdogan@ku.edu.tr

# Abstract

The current methods for characterization of frequency-dependent material properties of human liver are very limited. In fact, there is almost no data available in the literature showing the variation in dynamic elastic modulus of healthy or diseased human liver as a function of excitation frequency. We show that frequency-dependent dynamic material properties of a whole human liver can be easily and efficiently characterized by an impact hammer. The procedure only involves a light impact force applied to the tested liver by a hand-held hammer. The results of our experiments conducted with 15 human livers harvested from the patients having some form of liver disease show that the proposed approach can successfully differentiate the level of fibrosis in human liver. We found that the storage moduli of the livers having no fibrosis (F0) and that of the cirrhotic livers (F4) varied from 10 to 20 kPa and 20 to 50 kPa for the frequency range of 0 to 80 Hz, respectively.

# 1. Introduction

Accurate characterization of the mechanical properties of soft tissues is important for diagnosing medical pathologies and developing solutions for them. With the recent advances in technologies leading to the development of surgical simulators, medical robots, and computer-assisted surgical planning systems, this topic has gained even more importance. While strain and time-dependent material properties have been investigated extensively, less attention has been paid to the frequency-dependent dynamic material properties. However, the dynamic response of soft tissues to periodic or impact loading is important in many areas of biomechanics and biomedical engineering. For example, frequency-dependent mechanical properties play a crucial role when investigating the mechanisms of organ injury that result from high-speed impact such as car accident. The propagation speed and radius of the impact wave depends on the dynamic material properties of the organ. Similarly, when designing prosthetic devices for lower-body amputates, it is important to know how the soft tissue responds to the periodic impacts coming from the ground.

The current methods of acquiring the frequency-dependent material properties of soft tissues involve the use of mechanical indenters, rotational rheometers, and medical imaging techniques. To characterize the frequency-dependent viscoelastic material properties of soft tissues, the most common method is the dynamic loading test: small periodic strains at varying frequencies are applied to the tissue sample and the stress response is recorded. Because of the viscoelastic nature of soft tissues, two stress components, one in-phase with the applied strain and the other out-of-phase, can be obtained from the measurements at each frequency. As a result, two elastic moduli, one in-phase and the other out-of-phase, can be calculated by taking the ratio of the stress to the applied strain at each frequency. These two moduli form the so-called complex modulus with the in phase modulus, also known as the storage modulus, being its real part and the out-of-phase modulus, also known as the loss modulus, its imaginary part. The storage and loss moduli define the energy storage and dissipation capacity of the tested soft tissue, respectively.

In this study, we investigate the frequency-dependent viscoelastic material properties of human liver. Most of the existing dynamic data on liver in literature is for animal tissues, which has been acquired via the dynamic loading test. Liu and Bilston (2000) investigated the linear viscoelastic properties of bovine liver using a generalized Maxwell model and conducted three types of experiments a) shear strain sweep oscillation, b) shear stress relaxation, and c) shear oscillation. The shear stress and strain were calculated based on the torsional load. In strain sweep oscillation experiments, the liver tissue was subjected to a sinusoidal angular torsion at a fixed frequency of 1, 5, or 20 Hz using a strain controlled rheometer. The strain amplitudes were gradually increased from 0.06% to 1.5% while the storage and loss moduli of the bovine liver were measured. In stress relaxation experiment, sudden torsional shear strain was applied to liver tissue for 0.02 seconds and the shear relaxation modulus was measured over 3000 seconds. Finally, in shear oscillation experiments performed in the range of 0.006 to 20 Hz, the storage and loss moduli were measured again. The results show that the shear relaxation modulus of bovine liver reaches to steady state around 0.6 kPa. The results of the oscillatory shear experiments show that the storage modulus of bovine liver increases from 1 kPa to 6 kPa with increasing frequency while the loss modulus increases to a peak value of 1 kPa at about 1 Hz and then decreases to 0.4 kPa as the frequency reaches to 20 Hz. Kiss et al. (2004) performed in vitro experiments with canine liver tissue to characterize its viscoelastic response. They measured the storage and the loss moduli of the liver tissue for the frequencies ranging from 0.1 to 400 Hz by applying cyclic stimuli to the tissue. The resulting moduli spectra were then fitted to a modified Kelvin–Voigt model, which was called as the Kelvin–Voigt fractional derivative model (KVFD) by the authors. They show that there is an excellent agreement between the experimental data and the KVFD model; particularly at frequencies less than 100 Hz. Valtorta and Mazza (2005) developed a torsional resonator to characterize the dynamic material properties of bovine and porcine livers. By controlling the vibration amplitude, shear strains of less than 0.2% were induced in the tissue. The experiments were performed at different eigenfrequencies of the torsional oscillator and the complex shear moduli of bovine and porcine livers were characterized in the range of 1–10 kHz. The results of the in vitro experiments on porcine liver show that the magnitude of complex shear modulus varies between 5-50 kPa depending on whether the data collected from the external surface or the internal section of the liver (as reported by the authors, the former leads to considerably larger shear stiffness due to the presence of the stiff capsula). The shear modulus of the bovine liver is shown to vary between 15-30 kPa.

Compared to the animal studies, the number of studies investigating the frequency-dependent material properties of human liver and pathologies through biomechanical measurements are very limited. Most of the earlier studies have focused on the characterization of static and viscoelastic material properties of animal (Kerdok et al., 2006, Samur et al., 2007, Rosen et al., 2008, Hu et al., 2008, Gao and Desai, 2010) and human livers (Nava et al., 2008), but not their frequency-dependent properties. Saraf et al. (2007) investigated the dynamic response of human liver in hydrostatic compression and simple shear using the Kolsky bar technique at high strain rates ranging from 300 to 5000 $s^{-1}$. This technique involves the use of two elastic pistons with a disk-shaped material sample inserted between their ends. A pressure wave is generated by applying an impact at the free end of one of the pistons. By measuring the difference of vibrations at the extremities of the structure, the mechanical properties of the sample can be obtained. Using this approach, Saraf et al. (2007) measured the bulk and the shear moduli of human liver under dynamic loading as 280 kPa and 37-340 kPa (depending on the strain rate), respectively. Mazza et al. (2007) conducted in vivo and ex vivo experiments with ten human subjects having some liver pathology. Static mechanical properties of human liver were measured at multiple locations using an intra-operative aspiration device. Most of the tests were performed on diseased liver segments undergoing subsequent resection. Measurements were performed by the surgeon on the normally perfused liver in vivo and on the resected specimen ex vivo. The relationship between mechanical parameters and various pathologic conditions affecting the tissue samples was quantified. The fibrotic tissue is found to be three times stiffer than the normal tissue. Later, Nava et al. (2008) performed aspiration experiments on healthy human liver with the same device. They estimated the long term and instantaneous linear elastic modulus of human liver as 20 kPa and 60 kPa respectively.

In summary, there are limited number of studies and data on the frequency-dependent material properties of healthy and diseased human liver measured by either mechanical indenter or a rheometer. The latter is not even practical for collecting data from a whole organ. Currently, the level of fibrosis in a diseased liver is assessed by biopsy which is an invasive and painful procedure. Alternative medical imaging techniques based on transient ultrasound elastography, called FibroScan (Sandrin et al., 2003) and Magnetic Resonance Elastography, MRE, (Manduca et al., 2001) have been developed to quantify liver fibrosis non-invasively. In both approaches, the measurements are performed externally without having any direct contact with the actual liver tissue. As a result, there are still open questions about the accuracy and the validity of FibroScan and MRE. Moreover, the liver elasticity values reported in FibroScan and MRE studies are typically measured at a certain frequency rather than a range of frequencies.

Finally, most of the earlier studies on human or animal liver have utilized the dynamic loading test for the characterization of frequency–dependent material properties. However, one drawback of the dynamic loading test is that the measurements have to be made at each frequency within the range of frequencies of interest, which may not **be** practical for measuring, for example, live tissue properties of human liver in place since the time interval for data collection is typically limited due to the adverse effect of the subject's breathing on the measurements. Moreover, the collected data can be erroneous for the same reason. Faster and more practical methods are desired for the dynamical characterization of soft organ tissues

In this study, we utilize impact test to characterize the frequency-dependent material properties of human liver and its pathologies. The technique involves the use of a hand-held hammer to apply a light impact force on a pre-load mass placed on the organ surface (see Figure 1a). The major contributions of this study can be listed as follows:

- Most of the earlier studies on human and animal liver have investigated the strain-dependent hyperelastic or time-dependent viscoelastic material properties. There is very limited data available on the frequency-dependent material properties of healthy and fibrotic human liver characterized by mechanical devices. The main goal of this study is to fill the gap in this area.
- We are not aware of any earlier study utilizing an impact hammer for the measurement of the dynamic material properties of soft organ tissues. Impact test performed with an impact hammer enables one to collect data faster than the dynamic loading test for a range of frequencies. The whole test takes less than a minute and no harm is made to the organ during the measurements.

- We present a new approach based on impact test for the characterization of frequency-dependent material properties of a "whole" organ having a variable cross-sectional area and length. The commercial measurement devices and approaches for the same purpose are typically designed to work with small samples of known geometry in a laboratory environment. Hence, our approach can be potentially used in in-vivo experiments to determine the live material properties of an organ in place.

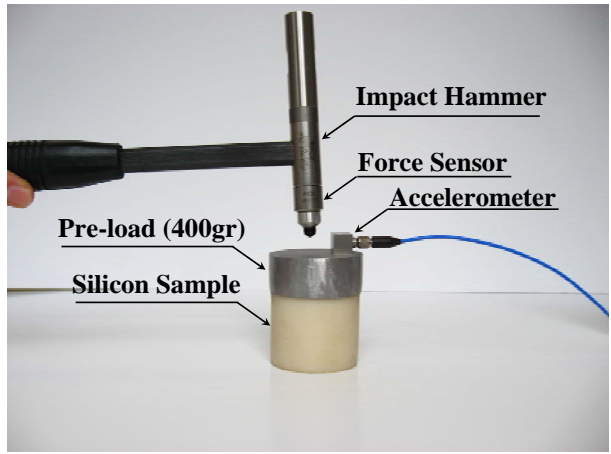

a)

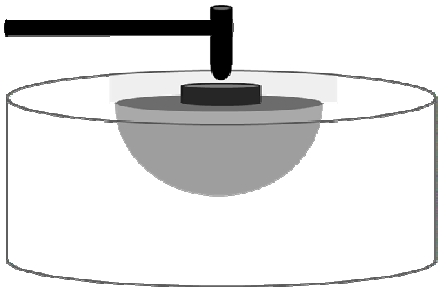

b)

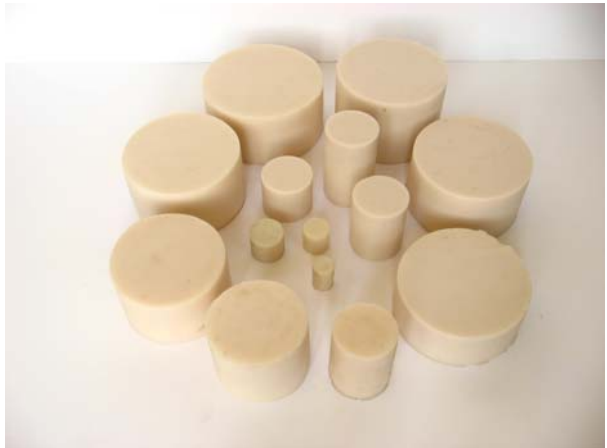

c)

**Figure 1**. a) The components of the measurement system. b) The propagation of impact load on a large sample. c) The family of silicon samples used in our experiments.

## 2. Methods

The impact test involves the use of a hand-held hammer to apply a light impact force on a pre-load mass placed on the test specimen (see Figure 1a). We model the dynamic response of the test specimen under impact loading using a hysteretic damping model as suggested by Nashif et al. (1985)

$$m_{eff}\ddot{x}(t) + k^* x(t) = f(t) \tag{1}$$

where $m_{eff}$ is the effective mass of the system including the mass of the pre-load placed on the top of the specimen and $f(t)$ is the impact force applied to the pre-load, which results in a displacement, $x(t)$. The complex dynamic stiffness of the specimen, $k^*$, is a function of the excitation frequency, $\omega$, and defined as $k^*(\omega) = k(\omega)(1 + \eta(\omega))$, where $k(\omega)$ is the dynamic stiffness and $\eta(\omega)$ is the loss factor. The above equation can be written in the frequency domain to obtain the frequency response function (FRF) as

$$T(j\omega) = \frac{X(j\omega)}{F(j\omega)} = \frac{1}{-m_{eff}\omega^2 + k(\omega)(1+\eta(\omega)\omega)} \tag{2}$$

Now, if we define $r$ as the ratio of the excitation frequency to the natural frequency of the system, $r = \omega/\omega_n$, then the dynamic stiffness and the loss factor of the specimen can be calculated from the measured FRF and the resonance frequency as suggested in Lin et al. (2005)

$$k(\omega) = \frac{\text{Re}(T(j\omega))}{|T(j\omega)|(1-r^2)}$$
$$\eta(\omega) = -\frac{\text{Im}(T(j\omega))}{\text{Re}(T(j\omega))}(1-r^2) \qquad (3)$$

After obtaining the dynamic stiffness, the dynamic elastic modulus, $E(\omega)$, can be calculated using the following relation

$$E(\omega) = \frac{k(\omega)L_{eff}}{A_{eff}} \qquad (4)$$

where, $L_{eff}$ and $A_{eff}$ are the effective length and cross sectional area of the specimen along the direction of the loading. In working with small specimens of known geometry, the effective length and cross-sectional area can be taken as the real values of the specimen in the calculations. Otherwise, as in the case of testing a whole organ, the effective values must be determined. However, it is not straightforward to calculate the effective cross-sectional area and length of a large specimen of varying size since the impact force applied to it propagates in a complex manner over its surface and along the direction of length, and is only effective within a radius of influence (see Figure 1b).

Once the dynamic elastic modulus is calculated, the complex elastic modulus is estimated as

$$E^*(\omega) = E(\omega)(1 + \eta(\omega)j) \qquad (5)$$

Then, the complex modulus $E^*(\omega)$ is written in terms of its real and imaginary parts as

$$E^*(\omega) = E_S(\omega) + E_L(\omega) j \qquad (6)$$

The real part, $E_S(\omega)$, is known as the storage modulus and an indicator of energy storage capacity of the material. The imaginary part, $E_L(\omega)$, is known as the loss modulus and related to the energy dissipation capacity of the material. The ratio of $E_L(\omega)/E_S(\omega)$ is equal to the loss factor $\eta(\omega)$ defined earlier.

## 3. Experiments

### 3.1. Impact Test

In our experiments, an impulse excitation force is applied to the cylindrical pre-load (radius = 25 mm, weight = 400 grams) placed on top of the specimen by the impact hammer (PCB Piezotronics Inc., Model 086C03, sensitivity is 2.1 mV/N) equipped with a force sensor (see Figure 1a). As suggested by the manufacturer, a soft tip and an extender mass were utilized for better response at low frequencies. The impulse response of the specimen was measured by a piezoelectric accelerometer (PCB Piezotronics Inc., Model 333B30, sensitivity is 101.2 mV/g, where g is the gravitational acceleration, range is 0.5-3000 Hz) attached to the pre-load using a thin film of adhesive wax. As suggested by the manufacturer, five measurements were taken from each test specimen and then the average values were used in the analysis. The accelerometer and the force sensor were connected to a dynamic signal analyzer (Data Physics Corporation, type SignalCalc Mobilyzer) for data processing. The FRF was obtained by taking the Fourier transform of the impulse response. In dynamic loading test, the same FRF is obtained by the frequency sweep method (i.e. small periodic strains are applied to the specimen and its force response is measured for a range of frequencies). Compared to the dynamic test, the impact test is more practical and the measurement time is much shorter.

Moreover, the results obtained by the impact test are as reliable as the ones obtained through the dynamic loading test (see Figure 2).

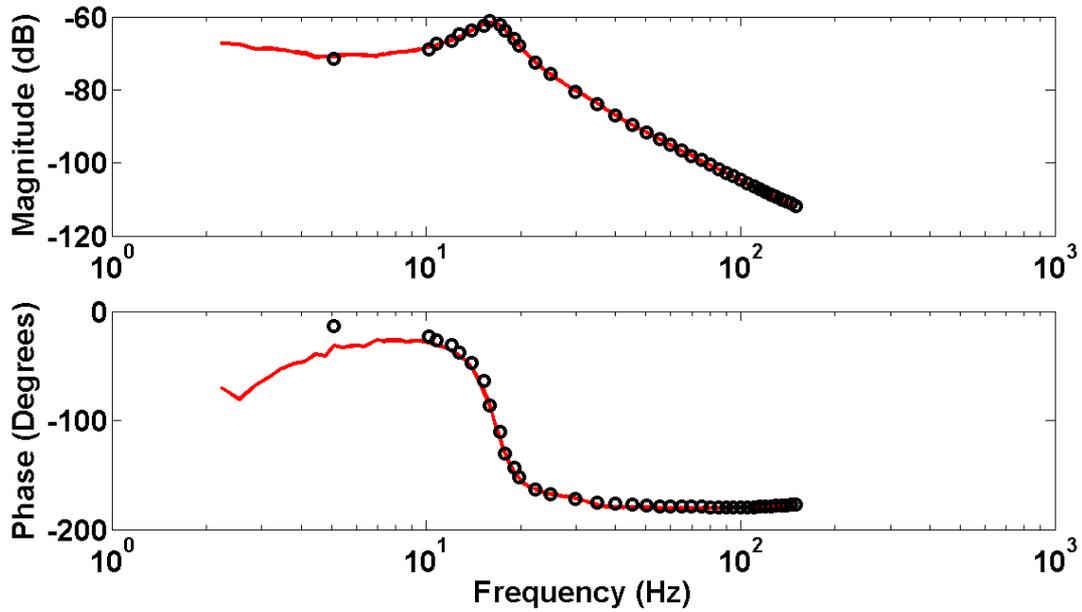

**Figure 2.** The data (dots) collected by a mechanical shaker from a cylindrical silicon sample via dynamic loading method is compared to the data collected by an impact hammer from the same sample (solid lines) in our laboratory.

**3.2. Experiments with Bovine Liver**

We first performed impact experiments in our laboratory with the "whole" bovine livers obtained from 2 different animals. The experiments were performed within 1 hour after the livers were harvested from the body. Figure 3 shows the results of the impact test performed on the whole liver of one of the animals. The dynamic stiffness, $k(\omega)$, and the loss factor, $\eta(\omega)$, of the whole liver (dashed lines) were calculated from the measured FRF as a function of excitation frequency using Eq. (3). One can observe the large variations in the dynamic stiffness and the loss factor around the natural frequency. As the excitation frequency approaches to the natural frequency ($r = 1$), the Eq. (3) becomes singular. **For this reason, the data around the natural frequency is not reliable.** To obtain a smooth curve for the dynamic stiffness and the loss factor, we fit a polynomial function (solid lines) to the experimental data, as suggested by Lin et al. (2005).

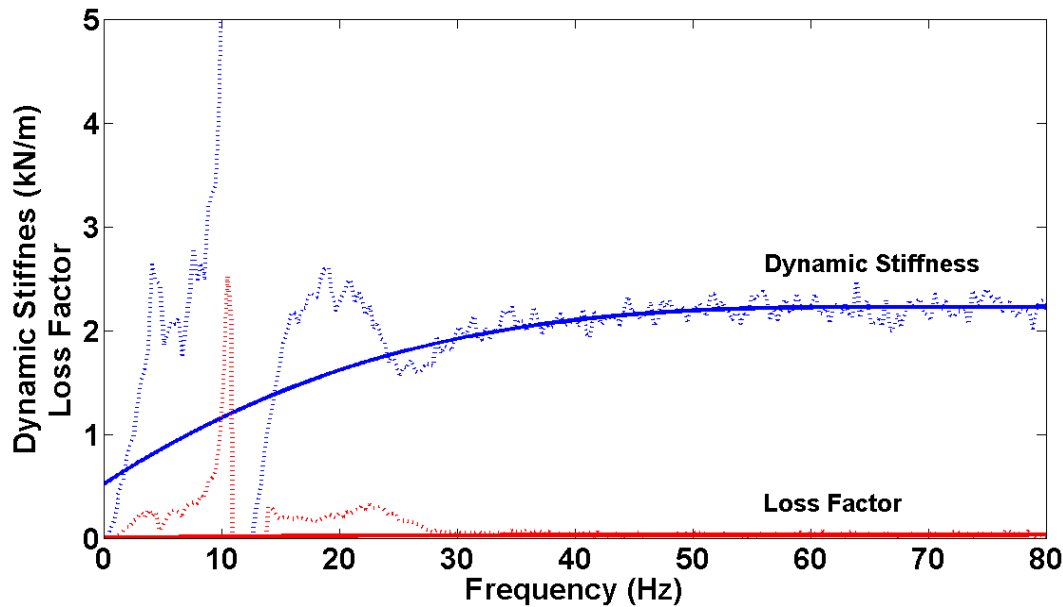

**Figure 3.** The dynamic stiffness and loss factor of a whole bovine liver.

It is important to emphasize that the dynamic stiffness, $k(\omega)$ depends on the sample geometry. However, the dynamic elastic modulus, $E(\omega)$, is a material property and independent of the sample size and shape. In order to estimate the dynamic elastic modulus of the bovine liver from the measurements taken on the whole organ, the effective length and cross-sectional area of the liver must be determined (see Eq. 4). Once the dynamic elastic modulus is determined, the storage and loss moduli can be easily obtained from it by using Eqs. (5) and (6). However, it is challenging to estimate the effective values of cross-sectional area and length of a whole liver at the measurement site. Otherwise, it is relatively straightforward to determine the dynamic material properties (storage and loss moduli) of the bovine liver if a small sample of known geometry is taken from the liver and the measurements are performed on this sample. Figure 4 shows the results of the material characterization experiments performed on the small cylindrical liver samples (radius = 23.25 mm, L = 60 mm) obtained from the animals. In the calculation of storage and loss moduli, the actual cross-sectional area and the length of the samples are used.

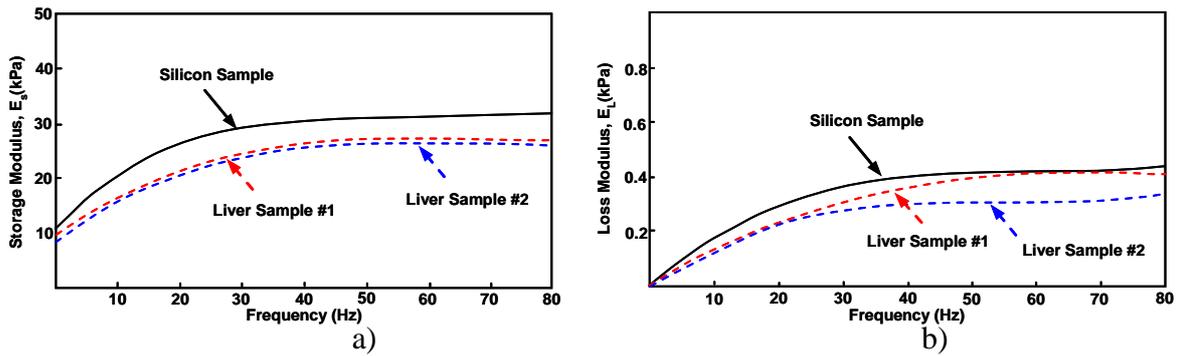

**Figure 4.** The storage (a) and loss moduli (b) of the bovine liver and the silicon samples in cylindrical shape.

### 3.3. Experiments with Silicon Samples

In order to determine the effect of length and cross-sectional area of the test specimen on its dynamic stiffness, we performed controlled experiments in our laboratory with cylindrical silicon samples having dynamic material properties (storage and loss moduli) similar to that of the bovine liver. Through these experiments, we determined the effective values of cross-sectional area and length that must be used for the whole bovine liver. We then compared the storage and loss moduli of the bovine liver estimated from the measurements performed on the whole bovine liver of unknown geometry with the one performed on the small cylindrical liver sample of known geometry, taken from the whole liver.

We prepared the silicon samples (see Figure 1c) using Smooth-Sil 910, which is frequently used in movie industry for modeling aliens, making masks, and replicating the human parts like arm, hand, and even face. Moreover, it can be easily obtained via shopping from internet or from a local distributor in many countries. Smooth-Sil 910 is a two-component silicone rubber: Part A is the base material that forms the product and Part B is the catalyst that hardens the silicon. For the softness adjustment, Silicon Oil was used. Silicon Oil does not change the structure of the silicon material but softens it. To prepare the samples in desired softness, Parts A and B and the oil were mixed in different ratios of mass. Frequent stirring was necessary for obtaining a homogenous mixture. Otherwise, air bubbles occurred in the mixture causing a difference in material properties locally. In order to minimize the stress concentration effects, the samples were molded in cylinder. As our molds, we used cylindrical glass pots for small size samples and PVC pipes for large size samples. The glass pots were preferred for the small samples since it was easy to get hardened silicon from it by breaking the glass. Several samples in different sizes were prepared for our experiments (see Figure 1c). Hence, the storage and loss moduli of all the silicon samples used in our experiments were similar to that of the bovine liver, but the dynamic stiffness of each sample was different depending on its length and cross-sectional area.

To investigate the effect of cross-sectional area on the dynamic stiffness of a sample, we kept the sample length constant (L = 60 mm) and varied the cross-sectional area of the samples with respect to the cross-sectional area of the pre-load. Figure 5a shows the dynamic stiffness curves of the silicon samples having different cross-sectional areas. The area ratio, AR, is defined as the ratio of the cross-sectional area of the cylindrical samples to the cross-sectional area of the pre-load. As the cross-sectional area of the sample is increased, its dynamic stiffness increases slowly at each frequency, reaching to a steady state value after AR > 5

(observe that the stiffness values of the samples for AR = 5.1 and AR = 6.7 almost overlap in Figure 5a). Figure 5b shows the changes in the resonance frequency of the same silicon samples as a function of AR. As shown in here again, the resonance frequency of the samples increases slowly as the cross-sectional area of the samples is increased, but reaches to a steady state value after a while. The solid line was obtained by curve-fitting an exponential function to the experimental data. The threshold value for AR was calculated accurately from this curve as $AR_{threshold}$ = 5.1 with less than 10% relative error. Hence, we conclude that, in testing silicon samples having cross-sectional area larger than the pre-load mass used in our experiments, the impact load is only effective within an area that is 5.1 times larger than the cross-sectional area of the pre-load.

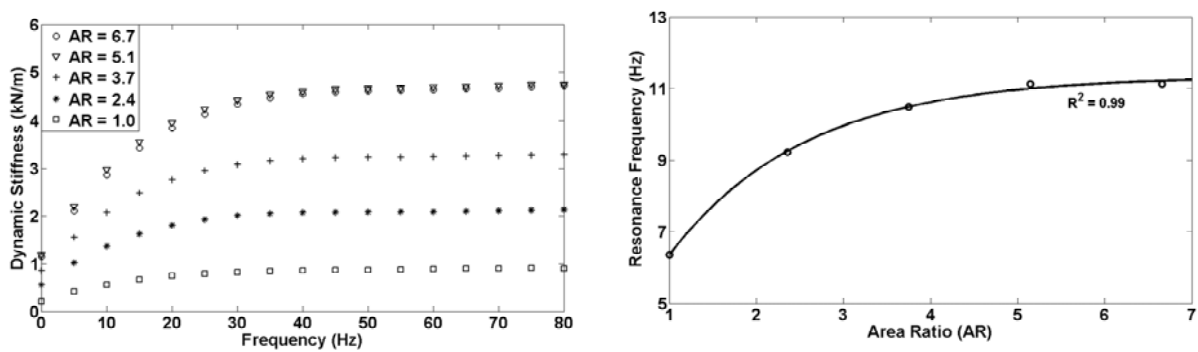

**Figure 5.** a) The variation in the dynamic stiffness of the silicon samples as a function of the cross-sectional area of the samples and the excitation frequency. b) The variation in the resonance frequency of the same samples as a function of AR.

To investigate the effect of the sample length on its dynamic stiffness, we kept the area ratio constant (a threshold value of AR = 5.1 is used to imitate the geometry of a whole liver) and varied the length of the samples. Figure 6a shows the variation in the dynamic stiffness of the samples for different sample lengths. As the length of the sample increases, the dynamic stiffness decreases slowly at each frequency. Figure 6b shows the changes in the resonance frequency of the same silicon samples as a function of the sample length. As shown in here again, the resonance frequency decreases as the sample length is increased, reaching to a steady value after L > 200 mm. The solid line is obtained by curve-fitting an exponential function to the experimental data. The threshold value for the length was calculated accurately from this curve as $L_{threshold}$ = 210 mm with less than 10% relative error. Hence, we conclude that, in testing large samples, the impact load is effective up to 210 mm only. In Figure 6c, we show the effect of pre-load mass on the storage modulus of a large sample (AR = 5.1). While the pre-load mass has some influence at low frequencies, it has almost no effect at high frequencies.

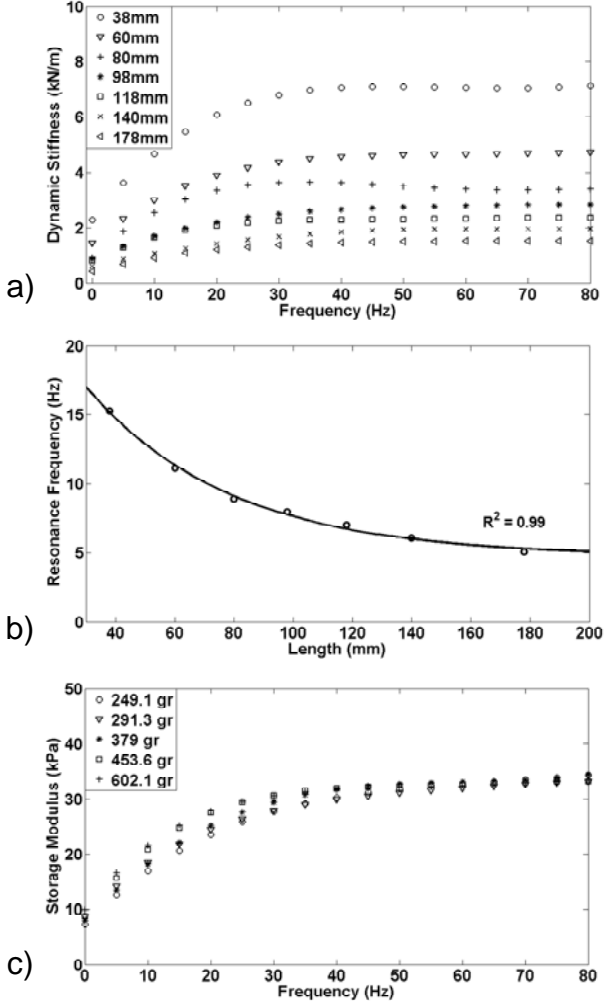

**Figure 6.** a) The variation in the dynamic stiffness of the large silicon samples as a function of the sample length and the excitation frequency. b) The variation in the resonance frequency of the same samples as a function of the sample length. c) The effect of pre-load mass on the storage modulus of a large sample.

To test the validity of the estimated values of effective cross-sectional area and the length, we compared the storage and loss moduli of bovine liver estimated from the whole organ of unknown geometry with the ones obtained from the small cylindrical liver samples of known geometry (see Figure 7). The results show a good agreement. In estimating the storage and loss moduli of bovine liver using the data collected from the whole liver (livers #1 and #2), the effective cross-sectional areas were taken as the 5.1 times the cross-sectional area of the pre-load (note that the cross-sectional areas of the bovine livers at the measurement site were significantly larger than that of the preload) and the effective lengths were taken as the actual length of each liver at the measurement location since they were significantly less than the threshold value of 210 mm. In the estimation of storage and loss moduli of the bovine liver using the data collected from the small cylindrical bovine samples (samples #1 and #2), no adjustment was necessary (i.e. the actual values of the cross-sectional area and the length of the samples were used in the calculations).

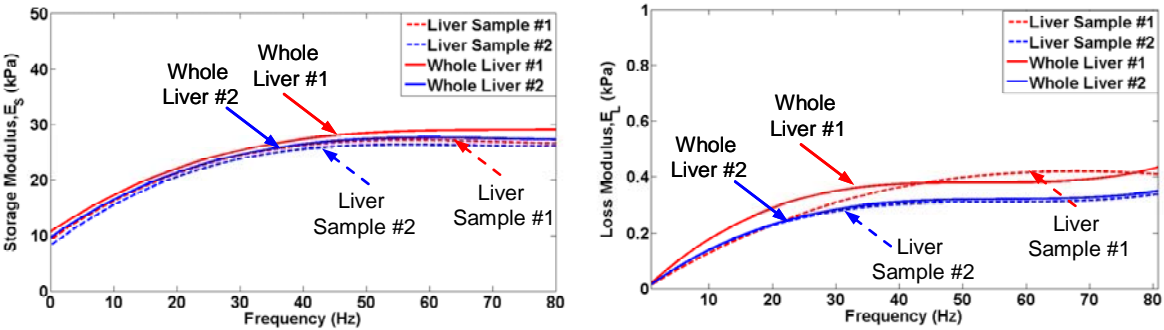

**Figure 7.** Comparison of the storage (a) and loss moduli (b) of bovine liver estimated from the whole organ of unknown geometry (solid lines) with the cylindrical liver samples of known geometry obtained from the same organ (dashed lines).

Using the experience and knowledge that we gained from the experiments performed with two bovine livers and several silicon samples, we finally performed impact experiments with human livers harvested from the liver patients having some level of fibrosis.

### 3.4. Human Experiments

A total of 15 liver samples were collected from the patients having a liver disease (age = 51 ± 10, gender = 12 male, 3 female). The study was approved by the ethical commission of the Florence Nightingale Hospital at Istanbul. All the tests were performed on the freshly excised livers. For each subject, the experimental data was collected at the operating room within 30 minutes after the liver was removed from the body. The length of the livers along the direction of impact loading was L = 81 ± 30 mm. The surface area of each excised liver was estimated from its digital image taken by a camera using a calibration grid (Figure 8). The height of each liver was measured using a digital micrometer at the location where the data was collected. The area ratio of livers was AR = 23.52 ± 7.06.

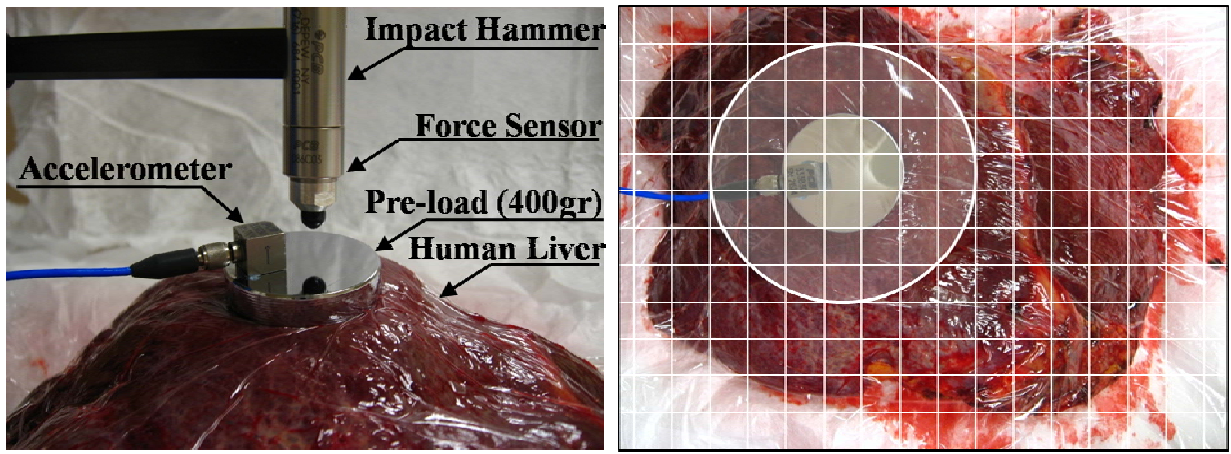

**Figure 8.** a) The dynamic material properties of human liver were measured as a function of frequency by an impact hammer. b) The impact load in our set-up is affective only within the surface area of liver (large white circle) that is 5.1 times greater than the cross-sectional area of the pre-load (small circle) placed on top of the liver.

Following the impact experiments at the operating room, each liver was sent to the Pathology Department of Istanbul University for histological examination. All livers were analyzed independently by an experienced pathologist blinded to the clinical data and the results of the impact tests. Fibrosis scores were assigned to each liver according to the METAVIR scoring system ranging from F0 to F4 (F0: no fibrosis, F1: portal fibrosis without septa, F2: septa fibrosis, F3: numerous septa without cirrhosis, and F4: cirrhosis). There were 2 patients in F0, 4 patients in F2, 4 patients in F3, and 5 patients in F4.

## 4. Results

Since the cross-sectional areas of all the human livers tested in our study were significantly larger than that of the pre-load used in our experiments, an effective value for the cross-sectional area was utilized in Eq. 4 to estimate the dynamic elastic modulus and subsequently the storage and loss moduli. This value was set to 5.1 times the cross-sectional area of the pre-load based on the results of the experiments performed with the tissue-like silicon samples. For the effective length, no adjustment was necessary since the lengths of all human livers at the measurement site were significantly less than the threshold value (210 mm). Figure 9 shows the results of the impact experiments performed on the human livers. The results show that the liver tissue becomes stiffer as the fibrosis score increases. Moreover, independent of the fibrosis level, the storage and loss moduli increase with an increase in excitation frequency. The storage modulus of the human livers with no fibrosis (F0) increased from 10 to 20 kPa as the excitation frequency is increased from 0 to 80 Hz and from 20 to 50 kPa for the livers with some level of fibrosis. The loss modulus of the human livers increased from 0 to 5 kPa with an increase in the excitation frequency from 0 to 80 Hz (Figure 9).

The storage and loss moduli of the human livers were compared for different levels of fibrosis using a two-way ANOVA and Tukey pair-wise comparisons. Two-way ANOVA showed significant effect of excitation frequency on the storage modulus, $F_{20,231} = 14.7$, $p < 0.001$, significant effect of fibrosis on the storage modulus, $F_{3,231} = 328.6$, $p < 0.001$, and no interaction between two, $F_{60,231} = 0.05$, $p = 1.0$. Subsequently, pair-wise comparisons were performed to compare the storage modulus of each fibrosis group with the others. The results showed significant differences in the storage modulus of patient groups having different levels of fibrosis (all p values were less than 0.001).

A similar analysis was performed for the loss modulus. Two-way ANOVA performed on the loss modulus showed significant effect of excitation frequency on the loss modulus, $F_{20,231} = 11.9$, $p < 0.001$, significant effect of fibrosis on the loss modulus, $F_{3,231} = 6.5$, $p < 0.001$, and no interaction between two, $F_{60,231} = 0.05$, $p = 1.0$. However, multiple pair-wise comparisons did not show significant differences between the groups except for the F2-F3 pair ($p < 0.001$).

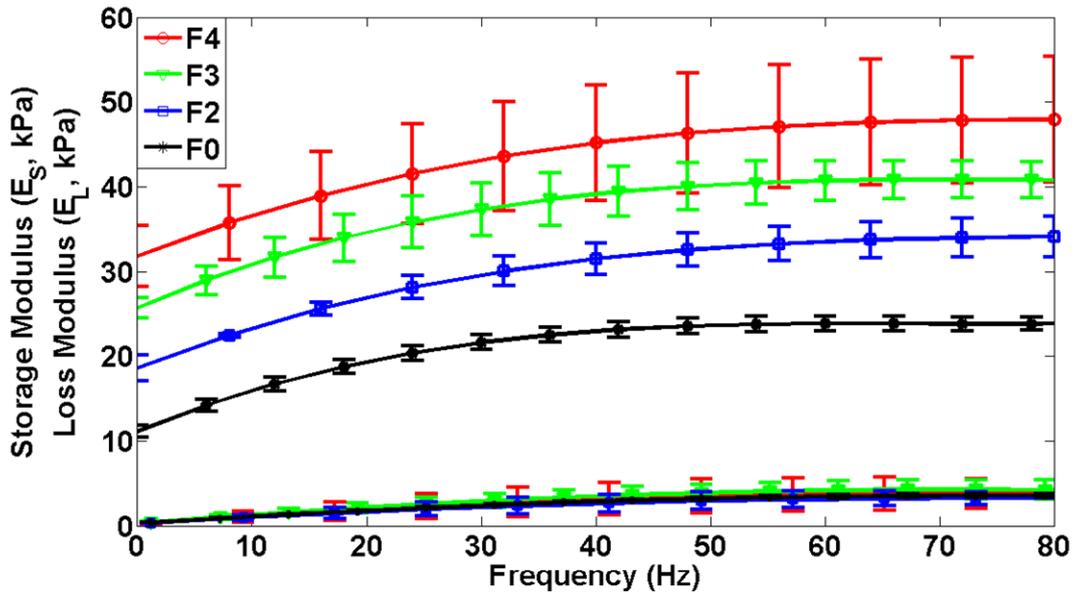

**Figure 9.** The storage (upper curves) and loss (lower curves) moduli of human liver as a function of frequency for different levels of fibrosis.

## 5. Discussion and Conclusion

We presented a new approach for the dynamic characterization of soft organ tissues using an impact hammer. Our approach is the extension of the experimental technique proposed by Lin et al. (2005) to measure hysteretic damping of rubber mounts. The original technique enables the measurement of the dynamic stiffness of small samples of known geometry only. If the cross-sectional area and the length of the sample are known, the dynamic elastic modulus can be easily estimated from the measured dynamic stiffness using Eq. (4). However, effective values of cross-sectional area and length are required to estimate the dynamic elastic modulus of an organ tissue from the impact experiments performed on the whole organ. For this reason, we investigated the effect of sample length and cross-sectional area on the dynamic stiffness of a viscoelastic material through a set of controlled experiments conducted with tissue-like silicon samples in our laboratory. We then performed ex-vivo experiments in an operating room with freshly harvested human livers to determine their storage and loss moduli using the effective cross-sectional area and the length values estimated from the tissue-like silicon experiments. Here, we assumed that the threshold values estimated for the cross-sectional area and length through the silicon experiments performed on large size silicon samples applies to the whole human liver. We tested this assumption by conducting experiments with two bovine livers. We compared the storage and loss moduli of the bovine liver estimated by the experimental data collected from two whole livers of unknown geometries with that of the cylindrical samples taken from the same livers (Figure 7). The results showed a good agreement. This encouraged us to use the same threshold values in the characterization of dynamic material properties of human liver. Ideally, it would be better if we had taken cylindrical samples from the harvested human livers and compare our results for the whole human livers with the ones obtained from these samples. However, the experimental protocol approved by the ethical committee of the hospital, where we performed the impact experiments, did not allow us to cut and take samples from the harvested livers of the patients. These livers were transferred immediately to the pathology department for the histological examination after the surgery and the impact experiments performed in the operating room. Alternatively, a finite element analysis could be conducted to investigate the effect of sample dimensions on the dynamic material properties of the liver.

As emphasized in the introduction section, there is limited number of biomechanical studies available in the literature on the dynamic material properties of healthy or diseased human liver to compare our results directly (see the next paragraph for an extensive discussion on the existing medical imaging studies in the same area). The magnitude of complex shear modulus values (5-50 kPa) reported in Valtorta and Mazza (2005) for porcine liver are comparable to the complex elastic modulus values reported for the human livers tested in our study. The Young modulus estimated for pig liver up to 4 mm indentation in the earlier in-vivo or in-situ studies (Ottensmeyer 2001, Samur et al., 2007, Hu et al., 2008) is around 10 kPa and comparable to the value that we estimated at zero frequency for the human liver having no fibrosis (Figure 9). In our current study, we found that the cirrhotic livers (F4) are more than twice stiffer than the livers having no fibrosis (see Figure 9). Mazza et al. (2007) reported that the fibrotic liver tissue is three times stiffer than the healthy liver tissue.

In particular, the evaluation of liver fibrosis is of great clinical interest. The early detection of liver fibrosis may lead to a more successful treatment. In fact, some cases of early fibrosis may be reversible with the elimination of the cause. Detecting the liver disease at the stage of cirrhosis, which is the end-stage progression of fibrosis, is relatively easy, but not much helpful from a clinical point of view. It is often more difficult to tell by clinical parameters how much fibrosis exists in a liver. Currently, the level of fibrosis in a diseased liver is assessed by biopsy which is an invasive and painful procedure. An alternative technique based on transient ultrasound elastography, called FibroScan (EchoSens S.A., Paris, France), has been developed by Sandrin et al. (2003) to quantify hepatic fibrosis non-invasively. This system is equipped with a probe consisting in an ultrasonic transducer mounted on the axis of a vibrator. A vibration of mild amplitude and low frequency (50 Hz) is transmitted from the vibrator to the tissue by the transducer itself. This vibration induces an elastic shear wave that propagates through the tissue. The propagation velocity of the wave within the tissue is directly related to the tissue stiffness (or elastic modulus). The harder the tissue, the faster the shear wave propagates. The results obtained by Sandrin et al. (2003) indicate that liver gets harder as fibrosis spreads out. They found that the elasticity of fibrotic livers varied from 3.35 to 69.1 kPa, where the largest variation was observed for F4 (14.4 to 69.1 kPa). In a more comprehensive study, Ziol et al., (2005) correlated the stiffness values of liver samples measured by FibroScan to the fibrosis stage. The elastic modulus of fibrotic liver was found to vary from 4.1 to 7.1 kPa for F0-1, from 4.8 to 9.6 kPa for F2, from 7.6 to 12.9 kPa for F3, and 16.3 to 48 kPa for F4. The cut-off values for deciding on the fibrosis level were determined as 8.8 kPa for $F \geq 2$, 9.6 kPa for $F \geq 3$, and 14.6 kPa for $F = 4$. Magnetic Resonance Elastography (MRE) is another noninvasive imaging technique that can be used to measure the stiffness of liver (Manduca et al., 2001, Rouviere et al., 2006). MRE is performed by transmitting shear waves within the liver and then imaging the waves using Magnetic Resonance techniques. Recent studies show that the MRE technique is also a feasible method to stage liver fibrosis and diagnose cirrhosis (Huwart et al., 2006). The results showed that the liver elasticity estimated by MRE increased with increasing stage of fibrosis. The shear modulus was found to fluctuate between ~2-7 kPa at the excitation frequency of 65 Hz, depending on the fibrosis level.

Our measurements are more compatible with the FibroScan measurements in terms of the range of elasticity values. However, a direct comparison of our results with the results obtained through the medical imaging studies is not possible due to the significant differences between the measurement approaches. First of all, both medical imaging techniques discussed above are external. Moreover the elasticity values reported in FibroScan and MRE studies are typically measured at a certain frequency rather than a range of frequencies. Finally, there are still open questions about the accuracy and the validity of both FibroScan and MRE. MRE has a potential but it is still in clinical research. FibroScan is commercially available, but this technique works best for separating patients with minimal or no fibrosis from those with significant fibrosis (Lucidarme, et al., 2009). Moreover, Bensamon et al. (2008) demonstrates the sensitivity of the FibroScan technique to local measurements. They compare the liver stiffness measured by FibroScan with MRE and finds higher variability in FibroScan measurements. This variation was attributed to the thickness of subcutaneous tissue and the movements of the probe during the exam, which effects the direction of the wave pressure and consequently the liver stiffness measurement.